\documentclass[11pt]{article}
\pdfoutput=1
\usepackage{amsmath,amsthm}
\usepackage{amssymb}
\usepackage{graphicx,mathrsfs}
\usepackage{geometry}       
\usepackage{dcolumn}   % needed for some tables
\usepackage{bm}        % for math
\usepackage{graphicx,mathrsfs}
\usepackage{nicefrac}
\usepackage{multirow}
\usepackage{color}

\def\ie{{\it i.e.}}

\newcommand{\be}{\begin{equation}}  
\newcommand{\ee}{\end{equation}}  
\newcommand{\bea}{\begin{eqnarray}}  
\newcommand{\eea}{\end{eqnarray}}  

\newcommand{\tr}{\operatorname{tr}}

\begin{document}

\vspace*{1.2cm}

\begin{center}

\thispagestyle{empty}
{\Large\bf 
%Taming systematic uncertainties at the LHC: \\
% Simplified likelihoods from the central limit theorem
Taming systematic uncertainties at the LHC with  the central limit theorem
}\\[10mm]

\renewcommand{\thefootnote}{\fnsymbol{footnote}}

{\large  Sylvain~Fichet  
\footnote{sylvain@ift.unesp.br} }\\[10mm]

\addtocounter{footnote}{-1}

{\it  ICTP South American Institute for Fundamental Research, Instituto de Fisica Teorica,\\
Sao Paulo State University, Brazil \\
}

\vspace*{12mm}

{  \bf  Abstract }
\end{center}

We study the simplifications occurring in any likelihood function in the presence of a large number of small systematic uncertainties.
We find that the marginalisation of these uncertainties can be done analytically by means of second-order error propagation, error  combination, the Lyapunov central limit theorem, and under  mild approximations which are typically satisfied for LHC likelihoods. 
The outcomes of this analysis are \textit{i}) a very light treatment of systematic uncertainties  \textit{ii}) a convenient  way of reporting the main effects of systematic  uncertainties, such as the detector effects occuring in LHC measurements.

\noindent
\clearpage

\section{Introduction}

%The search for  physics beyond the Standard Model at a collider consists in observing and analysing random events.

The search for physics beyond the Standard Model  requires a thorough statistical investigation  of the  data collected at the LHC.  
The analyses of LHC samples are often plagued by a number of systematic uncertainties, such as detector resolutions or the imperfect knowledge of physical constants,  that should be treated with care in order to obtain reliable conclusions on a putative  signal.
 A consistent implementation of these systematic uncertainties in the likelihood function is  done either by summing over all the realisations of the nuisance parameters or  by maximising the likelihood with respect to them,  respectively within the Bayesian and frequentist frameworks.

While the operations of Bayesian and frequentist marginalisation are conceptually clear, one can  distinguish two main issues related to their practical implementation. First, the correct treatment of systematic uncertainties  is technically challenging to perform, because it necessitates many multidimensional integrations or maximisations. For example, integrating the $O(4000)$ nuisance parameters of the Higgs likelihood is very difficult, even with the computing power accessible to the LHC Collaborations.
Second, transmitting or making public the information of the systematic uncertainties can also be technically challenging.  
The study we carry out will end up providing new insights into both of these topics.

In the context of the LHC,  communicating the full experimental likelihoods via the  \texttt{RooFit/Roostats} framework \cite{RooFit,Roostats} has been suggested in \cite{PresNP1,PresNP2}. The presentation method we will propose is somehow complementary from the proposal of \cite{PresNP1,PresNP2}, in that it is technically straightforward to carry out and leads to a fairly human-readable summary of the systematic uncertainties. Also, the goal of presenting LHC results decoupled from systematic uncertainties has been pursued  in \cite{Cranmer:2013hia} in the context of theoretical errors on Higgs cross-sections. While the objectives of \cite{Cranmer:2013hia} are partly similar to the ones of this work, the results obtained are different. In particular, the general marginal likelihood that we will display is derived from first principles, and  no discussion about  reparametrisation templates is required.

Along this work we are going to adopt the hypothesis of a \textit{large number of small independent sources of uncertainties}. 
Here by ``small'' we mean to qualify the \textit{relative magnitude} of the systematic uncertainties.  
 We stress that this is an intrinsic property that  does not depend on the magnitude of the  statistical uncertainties, \ie~on the size of the data sample.  The validity of our results  will thus not depend on the amount of observed data. 

The assumption of small relative magnitude  is fairly weak as, to our knowledge,  most of LHC systematic uncertainties have a relative magnitude which is much lower than $100\%$.
%This will be used for error propagation. 
The assumption of independence follows naturally from the process of describing systematic uncertainty as correctly as possible, as the more one delves into the origin of uncertainty, the more its description becomes a set of elementary sources   unrelated to each other.
Independence will have the crucial implication that the combination of the elementary  uncertainties 
 is  mostly described by its  \textit{first and second moments}. \footnote{This well-known fact is quantified by the Berry-Esseen theorem \cite{Berry,Esseen,Feller} and is closely related to central limit theorems.}

Our computation consists in two  steps of error propagation and error combination, that are laid down in Sec.~\ref{se:taming}. These   steps already lead  to a substantially simplified likelihood. In addition, if the relative magnitude of the combined uncertainties is  somewhat small with respect to one, we  show in Sec.~\ref{se:anal} that marginalisation can be done exactly, providing a general, explicit formulation of the marginal likelihood.  The cases of signal+background and differential distributions are treated.  Finally, a signal strength toy-analysis   illustrating the validity of our calculation is displayed in Sec.~\ref{se:example}.

\section{Taming a large number of small uncertainties}
\label{se:taming}

We consider an event-counting likelihood where $ \hat n$  is the observed event number and $ n$ is an expected (\ie~theoretical) event number. Assume $ n$ depends on parameters of interest $\theta$ and on nuisance parameters $\delta$. The likelihood is then defined as
\be
L(\theta,\delta ) \equiv 
{\rm Pr}\Big(\hat n\,\Big|\,n(\theta,\delta),\theta,\delta  \Big)\,.
\ee
All the variables written in this likelihood ($\theta,\delta,\hat n,n$)
should be understood as vectors, whose labels and dimensions will be made explicit below.

Without much loss of generality, \footnote{
Most of LHC event selections are independent from each other. A notable exception occurs when 
various overlapping selections of a same dataset are reported, with no information about more elementary, mutually exclusive selections. The correct statistics describing the set of overlapping selections is a multivariate Poisson. This  has been described in details in \cite{Fichet:2015yia} in the context of diboson ATLAS results \cite{Aad:2015owa}. 
} one further assumes that the various measurements of $n$ are statistically independent. In the following we will denote by $N$ the number of independent measurements, \ie~$I\in[1\ldots N]$, and by $p$ the number of nuisance parameters. The likelihood of our focus has thus the form
\be
L(\theta,\delta ) \equiv \prod_{I=1}^N
{\rm Pr}\Big(\hat n_I\,\Big|\,n_I(\theta,\delta),\theta, \delta  \Big)\,.
\label{eq:Likeind0}
\ee

The approach laid down in the present section applies in fact to any likelihood that can be expressed as a function of the expected event numbers, $L[n_I(\theta,\delta)]$.  \footnote{This includes the case where some systematic uncertainties have a large relative magnitude, and  have to be treated exactly instead of using the framework presented in this section. However, the best strategy in that case may be to first implement the small systematic uncertainties, leading to the approximate marginal likelihood of Eq.~\eqref{eq:Ltildegen}, then marginalise exactly (numerically) over the  large ones. Thus, even in that case, it is enough to start our analysis with the form Eq.~\ref{eq:Likeind0}.
 }
However, the subsequent analytical marginalisation presented in Sec.~\ref{se:anal} would not hold in general, as the approximate likelihood would not be Poissonian or Gaussian in $\delta$.

\subsection{Parametrisation}

For a  systematic uncertainty spanning a given domain, there exists in principle an infinite number of parametrisations, that are all equivalent under suitable redefinition of the  distribution of the uncertainty. Among all possible  parametrisations, it is useful to choose one that makes appear the relative magnitude of the uncertainty. We  define   a standardised representation as follows.
 
 For any quantity $A$ subject to uncertainty, that can take \textit{both signs}, one simply defines 
\be
A=A_0(1+\Delta \delta)\,,
\ee
where the nuisance parameter $\delta$   satisfies
\footnote{Given a random variable $X$ with density $f_X$, the expectation and variance operators are defined as 
${\rm E}(X)=\int dx\, x f_X(x)$, ${\rm V}(X)=\int dx \,x^2f_X(x)-{\rm E}(X)^2$.
 }

 \be{\rm E}[\delta]=0\,,\quad {\rm V}[\delta]=1\,, \label{eq:defdelta} \ee 
so that $\Delta$ corresponds to the relative magnitude of the uncertainty, \ie~ ${\rm V}[A]=A_0^2 \Delta^2  $.  Our working hypothesis of small relative magnitude translates as $\Delta\ll 1$. %This form can be seen as a linear expansion of any other form involving explicitly  $\Delta$, $A=A_0 f(\Delta\delta)$.

We also need to consider quantities that can only be positive, in the first place the expected event number $n$.
\footnote{ The linear parametrisation Eq.~\eqref{eq:defdelta} is not so suitable in that case because it requires to truncate the domain of $\delta$ above $ -\frac{1}{\Delta}$, implying that the prior itself depends on $\Delta$. 
%Moreover, if the distribution of $\delta$ do not drop fast enough near $ -\frac{1}{\Delta}$, \ie favours too much  there are cases where 
%the marginal likelihood can blow up for large values 
%
%, because, while even though $\delta$ is typically $O(1)$, it can in principle reach arbitrary large  negative values. This potentially introduces inconsistencies further in the analysis, even if one truncates the prior to ensure $\delta > -\frac{1}{\Delta}$. 
}
Throughout this paper one defines the standardised form for the error on a positive quantity as
\be
n\equiv n_0 e^{\delta \Delta}\,, \label{eq:deltaexp}
\ee
where ${\rm E}[\delta]=0$, ${\rm V}[\delta]=1$ and $n_0$ is the nominal value of $n$ in the absence of uncertainty. 
  Below it will become clear that the expansion in $\Delta$ has to be done up to  second order, so that Eq.~\eqref{eq:deltaexp} can be equivalently taken to be 
\be n\equiv n_0 \left(1+\delta \Delta+\frac{(\delta \Delta)^2}{2}+O(\Delta^3)\right)\,. \ee Compared to the linear form Eq.~\eqref{eq:defdelta}, one can see that the extra quadratic term ensures positivity of $n$. It also induces a small, positive shift of the mean value of $n$, as ${\rm E}(n)=n_0(1+\Delta^2/2)$ -- or similarly ${\rm E}(n)=n_0 e^{\Delta^2/2}$ without the expansion. The variance is ${\rm V}(n)=n_0^2(e^{2\Delta^2}-e^{\Delta^2})=n_0^2\Delta^2(1 +O(\Delta^4))$.

\subsection{Error propagation}

As a first step, we want to propagate the systematic uncertainties at the level of the event numbers.  For an event number $n$ depending on a quantity $Q$ subject to uncertainty, we have \be n[Q]\equiv n[Q_0(1+\Delta_Q \delta)]\,.\ee The propagation  amounts to perform a Taylor expansion with respect to $\Delta_Q $. This expansion should be truncated appropriately  to retain the leading effects of the systematic uncertainties in the likelihood.
 For now we take for granted that the expansion should be truncated above \textit{second order}. This order will be justified further below. 
 
In a one-parameter case, second order propagation leads to 
\be n\equiv n_0\exp\left( \frac{n'}{n_0} \Delta_Q \, \delta+
\left(\frac{n''}{n_0}-\frac{n'^2}{n_0^2}\right) \frac{\Delta_Q^2 \, \delta^2}{2} 
+O\left(\frac{n^{(3)}}{n_0}\Delta_Q^3\right)
\right)\,.
\label{eq:prop}
\ee
This is most easily obtained by expanding $\log n$.
Clearly, the validity of this expansion relies on neglecting higher powers of $\Delta_Q$ times the appropriate derivative of $\log n$. As long as $n$ is well-behaved, which should be checked in practice, this expansion is valid for uncertainties  that have a small relative magnitude, \ie
\be
\Delta_Q \ll 1\,.
\ee
For example, for elementary systematic uncertainties that do not exceed $\Delta \sim 10\%$, keeping the expansion up to second degree implies that the neglected higher order terms are  $O(0.1\%)$.

In the case were various uncertainties $Q_{1\ldots p}$ are propagated into $n$, it is convenient to use a vector notation for $\delta$ and $\Delta$. Assuming $p$ nuisance parameters, one defines 
\be
\delta=(\delta_1\ldots \delta_p)^t\,, \label{eq:stddelta}
\ee
\be
\Delta_1=\left(\frac{\partial n }{\partial \delta_{1}}\Delta_{Q,1},\ldots,
\frac{ \partial n }{\partial \delta_{p}}\Delta_{Q,p} \right)_{\delta=0}^t n_0^{-1}\,, \label{eq:stdDelta1}
\ee
\be
\Delta_2=\left(\frac{\partial^2 n}{\partial \delta_{i}\partial \delta_{j}}\frac{\Delta_{Q,i}\Delta_{Q,j}}{2n_0}\,- \frac{\Delta_{1,i}\Delta_{1,j}}{2}  \right)_{\delta=0}\,. \label{eq:stdDelta2}
\ee
 The relative uncertainties propagated to $n$ are then written as
\be n\equiv n_0\exp\left(  \Delta_{1}^t\cdot \, \delta+\delta^t\cdot\Delta_{2} \cdot \delta +O\left(\frac{n^{(3)}}{n_0}\Delta_Q^3\right)
\right)\,.
\label{eq:propmult}
\ee
After this step of error propagation, the likelihood takes the form
\be
L(\theta,\delta ) \equiv \prod_I
{\rm Pr}\Big(\hat n_I\,\Big|\,n_{0,I}\exp(\Delta_{1,I}\cdot\delta+\delta\cdot\Delta_{2,I}\cdot\delta), \theta, \delta  \Big)\,.
\label{eq:Likeind}
\ee
All the $n_{0,I}$, $\Delta_{1,I}$, $\Delta_{2,I}$  depend in principle on the parameters of interest $\theta$.

\textit{Details about the order of truncation.}  For the sake of determining the truncation order, it is enough to consider a one-parameter case and take limits. Consider a likelihood ${\rm Pr}(\hat n| n_0\exp(\Delta_1\delta+\Delta_2\delta^2))$ with $\Delta_2=O(\Delta_1^2)$.

We first study the  limit of an infinite amount of data.  In that case, the likelihood tends to a Dirac peak, \footnote{More precisely, the Dirac limit can be taken when the relative magnitude of the statistical uncertainty -- given by $1/\sqrt{n_0}$ in the Poisson case -- is small with respect to the inverse Fisher information of all the priors present in the problem.}
\be
\boldsymbol{\delta}\big(\hat n -n_0\exp(\Delta_1\delta+\Delta_2\delta^2)\big)\,.
\ee 
If one neglects the $\Delta_2$ term, marginalising this likelihood with a prior $\pi(\delta)$ for the nuisance parameter gives \be \tilde L=\pi\left( \log (\hat n/n_0)/\Delta_1\right)\,. \ee As $\delta$ is  $O(1)$ by definition, it follows that $\log (\hat n/n_0)=O(\Delta_1)$. Using this fact, one can then verify that $\tilde L = \pi\big( \frac{\log (\hat n/n_0)}{ \Delta_1 } +O(\Delta_1)\big) $ once  the $\Delta_2$ term is included. We  conclude that the leading effect of the systematic uncertainty comes  from $\Delta_1$, and appears thus at first order in the expansion.

Second we study the case where the amount of data is small enough so that the likelihood itself can be expanded with respect  to $n_0\Delta_1$, $n_0\Delta_2$. It comes
\be
L(\delta)=L(0)+ \delta\left( n_0\Delta_1  L'\right)+ \delta^2\left(\frac{\Delta^2_1}{2}(n_0^2L''+n_0L')+n_0\Delta_2L'\right) +O(n_0^3\Delta_1^3)\,.
\ee
One then marginalises this likelihood with respect to the nuisance parameter $\delta$ using an arbitrary prior, $\int d\delta L(\delta) \pi(\delta) $. By definition, ${\rm E}(\delta)=0$ (see Eq.~\eqref{eq:defdelta}), so the linear term vanishes. The leading effect of the uncertainties appears thus from the second order term in the expansion. This implies that the expansion has to be done at quadratic order from the beginning, so that the $\Delta_2$ term should not be neglected. 

As the truncation has to be done above second order in one of the limiting cases, it is convenient to use this  order in all cases  to ensure that all leading effects of systematic uncertainties are consistently taken into account.

%
%The picture is therefore that error propagation must be done up to quadratic order, unless all statistical errors 
%
%When the statistical error is small (\textit{cf} second case above), error propagation at linear order is enough to contain the main effect of systematic uncertainties. As long as this assumption is not true, error propagation must go up to second order. In the following 

\subsection{Error combination}

The previous step of propagation opens up the possibility of combining the nuisance parameters. 
We  define  nuisances parameter $\bar\delta_I$,  associated to every measurement $I$, so that
\be
n_I= \bar n_{0,I}\exp(\Delta_I\bar \delta_I)
\equiv
 n_{0,I}\exp\left(\Delta_{1,I}\cdot \delta+\delta^t\cdot\Delta_{2,I}\cdot\delta\right)\,. \label{eq:comb}
\ee
These combined nuisance parameters are in general \textit{correlated}  with each others, their joint distribution we will denote $\bar \pi$.
The set of equations \eqref{eq:comb} is the starting point for the combination of uncertainties. 
 \footnote{
At the level of the likelihood, combination is defined as the variable change
\be
\int d\bar \delta \bar L(\theta,\bar \delta) \bar\pi(\bar \delta)\propto \int d\delta L(\theta,\delta) \prod_{i=1}^p \pi_i(\delta_i)\,, \label{eq:comb2}
\ee
where the $\pi_i$ are the priors of the elementary nuisance parameters.
This is equivalent to Eq.~\eqref{eq:comb}. %From both equations \eqref{eq:comb} and \eqref{eq:comb2}, one gets that the combined prior is given by the usual variable change formula \be\pi_I(\delta_I)\propto \int d\delta_i\, \pi_i(\delta_I-(\Delta_{1,I}\cdot\delta+\delta^t\cdot\Delta_{2,I})) \boldsymbol{\delta}(\delta_i)\,, \ee where $\boldsymbol{\delta}$ is the Dirac distribution.
}
The likelihood expressed with respect to the combined nuisance parameters is written as
\be
\bar L(\theta, \bar \delta ) \equiv \prod_I
{\rm Pr}\Big(\hat n_{I}\,\Big|\,\bar n_{0,I}\exp(\Delta_{I}\delta_I), \theta, \bar \delta  \Big)\,.
\label{eq:Likecomb}
\ee
 Following our conventions, the combined nuisance parameters have to satisfy ${\rm E}(\bar \delta_I)=\int d\bar \delta_I\bar \delta_I \bar\pi(\bar \delta_I)=0$, ${\rm V}(\bar \delta_I)=\int d\bar \delta_I\bar \delta_I^2 \bar\pi(\bar \delta_I)=1$. The next task is to determine the numbers $\bar n_{0,I}$ and $\Delta_I$. This is obtained by taking the expectation and the variance on both sides of Eq.~\eqref{eq:comb}.
 
The central value of the event numbers before and after combination are different because of the nonlinear propagation.
It turns out that the diagonal terms of $\Delta_{2,I}$ contribute to the mean value of $n_I$, 
so that 
\footnote{
This shift can also  be observed by evaluating the mean value of the likelihood taken as a function of the $n_{0,I}$.
It can be derived explicitly for a Poisson likelihood, and can also  be derived for an arbitrary likelihood in the $n_{0,I}\Delta_2\ll1$ case, by expanding in $n_{0,I}\Delta_2$ and integrating by part with respect to the $n_{0,I}$. 
%%The effect of the $\delta^2$ term on the likelihood  can be derived by computing the mean value of the likelihood. For a Poisson distribution this can be seen explicitly for any $n$. For $n\Delta_2$, one can derive the shift for any likelihood by expanding $L(\lambda(1+\Delta_2\delta^2))=L(\lambda)+\Delta_2\delta^2L'(\lambda)$ an use integration by part on the second term. It comes that $\int d\lambda\delta L(1+\Delta_2\delta^2))\pi(\delta)=\int d\lambda L(1+\Delta_2\delta^2))\pi(\delta)=1-\Delta$ and that $E(n_0)=$
 }
\be
\bar n_{0,I}=n_{0,I}\left(1+\tr(\Delta_{2,I})\right) \,. \label{eq:n0I}
\ee
The  relative magnitudes $\Delta_I$ are obtained by evaluating the variance on the two sides of Eq.~\eqref{eq:comb}. One gets 
\be
\Delta_{I}=\left(\Delta_{1,I}^t\cdot\Delta_{1,I}\right)^{1/2}+O(\Delta^4)\,. \label{eq:DeltaI}
\ee
where the $O(\Delta^4)$ denotes higher order terms like $\tr(\Delta_{2,I}^t\cdot\Delta_{2,I})$, $(\tr\Delta_{2,I})^2$, $\Delta_{1,I}^t\cdot\Delta_{1,I}\tr\Delta_{2,I}$.
One may note  the contrast of Eq.~\eqref{eq:DeltaI} with the mean value Eq.~\eqref{eq:n0I}, where  $\Delta_{2,I}$ provides the main correction and cannot be ignored.

The next step is to compute the correlation matrix among the event numbers  $n_I$, $n_J$ induced by the systematic uncertainties.  
 The correlation matrix is found to be \footnote{In this paper we focus on the case of independent sources of uncertainty. The more general case of correlated nuisance parameters will be treated in \cite{usWIP}.}
\be
\rho_{IJ}=\frac{\Delta_{1,I}^t\cdot\Delta_{1,J}}{\Delta_I\Delta_J}
+O(\Delta^2)
\label{eq:corr}\,.
\ee
%
%We can see that the leading effect of the second order correction $\Delta_2$ is  a shift of the expected central values $n_{0,I}$.  An important subtlety remains to be discussed about Eq.~\eqref{eq:n0I}. Indeed, 
%
%This approach is valid 
%
%
%At that point, we have obtained the covariance matrix of the nuisance parameters. 
  In the next paragraph it will be made clear that under our working assumptions, this information is enough  to describe the entire shape of the combined uncertainties distribution $\bar\pi$.

\subsection{Shape of the combined prior}

Computing the joint distribution $\bar\pi$ of the combined uncertainties may seem at first view a very challenging task, as   
there can be in principle a lot of uncertainty sources. The experimental Higgs likelihood, for example,  contains $O(4000)$ nuisance parameters, \ie~the vector  $\delta^{}$ has dimension $O(4000)$. This means that 4000 convolutions would have to be done for each value of the combined nuisance parameters $\bar \delta_I$.

One should however realise that the shape of $\bar \pi$ is determined by its central moments of order higher than two, which all depend only on the $\Delta_{1,I}$ at leading order in the $\Delta$-expansion. In fact,
at leading order,
  the $\Delta_2$ term matters only for the mean value  and always gives subleading contributions to higher moments. This can be seen by evaluating the central moments of Eq.~\eqref{eq:comb}.   One can thus safely neglect the $O(\Delta_2)$ term in  the combination $\Delta_I\bar \delta_I=\Delta_{1,I}\cdot \delta +O(\Delta_2)$, and make the crucial observation that this quantity is as sum of many independent random variables.

Besides, one notices that all the common distributions for nuisance parameters, such as the uniform, normal, log-normal distributions, possess \textit{finite} higher moments. This is enough to invoke the Lyapunov central limit theorem (CLT) \cite{Billingsley}, \footnote{The Lyapunov CLT does not require identically distributed nuisance parameters, nor identical variances. In a similar fashion, the Lindberg-Feller CLT \cite{ Feller,Billingsley,Lindeberg} applies with a condition weaker than Eq.~\eqref{eq:CLT_lyapu}, but maybe less intuitive. The Lindberg condition is implied by the Lyapunov condition. 
} which  can be stated as follows. If it exists an integer $\kappa>0$ so that
\be
\frac{1}{\left(\sum^n_{i=1} \Delta_{1,I,(i)}^{2} \right)^{1+\kappa/2}}
\sum_{i=1}^n\Delta_{1,I,(i)}^{2+\kappa}{\rm E}[\delta_{1,(i)}^{2+\kappa}  ]\rightarrow 0 \quad \textrm{when}\quad n\rightarrow \infty\,, \label{eq:CLT_lyapu}
\ee
then the distribution of the combination $\Delta_{1,I}\cdot \delta$ converges in distribution towards a normal law with variance $\Delta_I^2$. For $\kappa=1$, for example, the condition involves the third moments of the nuisance parameters. The Lyapunov condition is verified for any kind of prior shape used in LHC analyses, such as normal, log-normal or uniform distributions. \footnote{In particular, note that  in cases where the distribution of the nuisance parameters is symmetric,   Eq.~\eqref{eq:CLT_lyapu} is zero for odd $\kappa$ and any $n$, so the Lyapunov condition is automatically satisfied. }

An estimate of the rate of convergence of the combined prior towards the normal law is given by  the Berry-Esseen theorem \cite{Berry,Esseen,Feller}. For the combination of identical nuisance parameters the maximal difference between the combined prior and the Gaussian  decreases as $1/\sqrt{n}$. For a combination of arbitrary nuisance parameters, which is our focus, the Berry-Esseen theorem states that the maximal difference between the combined prior and the Gaussian  is of order 
\be
\frac{1}{\left(\sum^n_{i=1} \Delta_{1,I,(i)}^{2} \right)^{3/2}}
\sum_{i=1}^n\Delta_{1,I,(i)}^{3}{\rm E}[\delta_{1,I,(i)}^{3}]\,.
\ee
This can be used in order to get  an estimate of the convergence of the combined prior.

The arguments above can be applied separately to every combined nuisance parameter $\bar \delta_I$. However, whereas the elementary uncertainties are independent, the various $\bar \delta_I$ are correlated between each other -- the correlation matrix is given by Eq.~\eqref{eq:corr}. The proof that the distribution of the set of $\bar \delta_I$ converges towards a multivariate normal is obtained by decomposing $\rho_{IJ}$  as $\rho_{IJ}=A^t_{IK}A_{KJ}$. \footnote{This is allowed as $\rho_{IJ}$ is a real symmetric matrix.} Provided that the Lyapunov condition is satisfied for every $A_{IK}\delta_K$, one gets by definition a multivariate normal with diagonal correlation matrix. Applying the reverse transformation achieves to proove that the combined prior has asymptotically the form
\be
\bar\pi(\bar \delta)= \frac{1}{(2\pi)^{N/2}|\rho|^{1/2}} \exp\left(- \frac{1}{2} \sum_{IJ=1}^N \bar \delta_I\,(\rho)^{-1}_{IJ}\,\bar \delta_J \right)\,.
\label{eq:picomb}
\ee
As a consequence, 
the combined uncertainty on the expected event numbers $n_I$ asymptotically follows  a \textit{multivariate log-normal} distribution (see Eq.~\eqref{eq:Likecomb}).

Besides these robust arguments, one may also note that many of the elementary systematics uncertainties readily have a Gaussian prior, which further enhances the convergence rate of the combination. This is also true for the log-normal distribution in the limit of relative magnitude, in which case the log-normal is approximately Gaussian. The manifestation of the CLT in the case of theoretical uncertainties for Higgs production and decay rates has been explicitly observed in \cite{Fichet:2015xla}.

\subsection{Practical considerations}
\label{se:pract}

 We claimed above that under the assumption of a large number of small uncertainties, the shape of elementary priors does not matter and the shape of the combined prior is approximately Gaussian. All the information needed to treat the systematic uncertainties is in fact contained in the mean values  and the covariance matrix of the combined nuisance parameters, that  are obtained through the steps of propagation/combination described above.

Some practical conclusions can already be drawn. It turns out that the approximate treatment proposed above 
 only requires the knowledge of a finite set of numbers: 
\begin{itemize}
\item The magnitude of the elementary uncertainties, $\Delta_Q^{i}$, of dimension $p$.
\item The first derivative of the expected event numbers with respect to every nuisance parameters, \ie~ $\partial n_I/\partial \delta_{i}$, of dimension $N\times p$. 
\item The diagonal second derivative of the expected event numbers with respect to every nuisance parameters, \ie~ $\partial^2 n_I/\partial \delta_{i}\partial \delta_{i}$, of dimension $N\times p$. 
\end{itemize}
All the relevant information about systematic uncertainties is thus encoded into  $(2N+1) p$ numbers.  The transmission of this information poses no technical challenge. In the context of LHC analyses, it could be an easy  and efficient way for the Collaborations of making public the main detector effects. 

 Besides, as a rule of thumb about the typical number of elementary uncertainties required for the CLT to converge,  one can ask for a minimum number of $p=4-5$   elementary uncertainties with similar magnitudes and flat priors. In case of Gaussian priors, this constraint does not hold as the combined prior is perfectly Gaussian for any $p$.

\section{Analytic marginalisation for Poisson and Gaussian likelihoods}
\label{se:anal}

The previous steps of propagation, combination, and prior simplification   can readily be used to  reduce the amount of nuisance parameters in any kind of fit. In the case of  Higgs experimental uncertainties, the $O(4000)$-dimensional space of nuisance parameters would be reduced to a $O(100)$-dimensional space -- the amount of statistically independently observed channels.
But ultimately, an integration still needs to be carried out over a space of substantially large dimension, for which a Monte Carlo integration is often required .

\begin{figure}
\centering
\includegraphics[scale=0.37,clip=true, trim= 0cm 0cm 0cm 0cm]{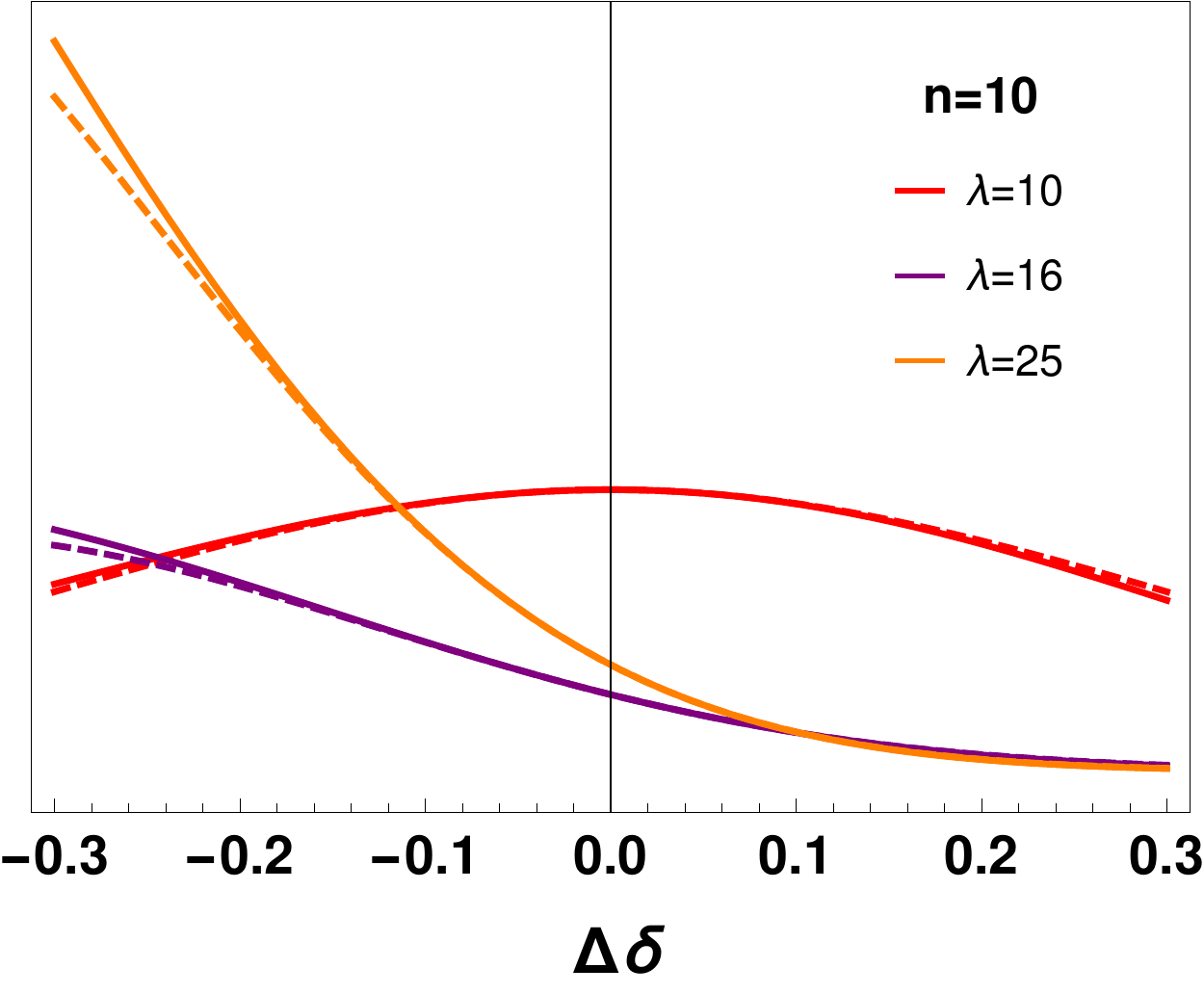}
\includegraphics[scale=0.37,clip=true, trim= 0cm 0cm 0cm 0cm]{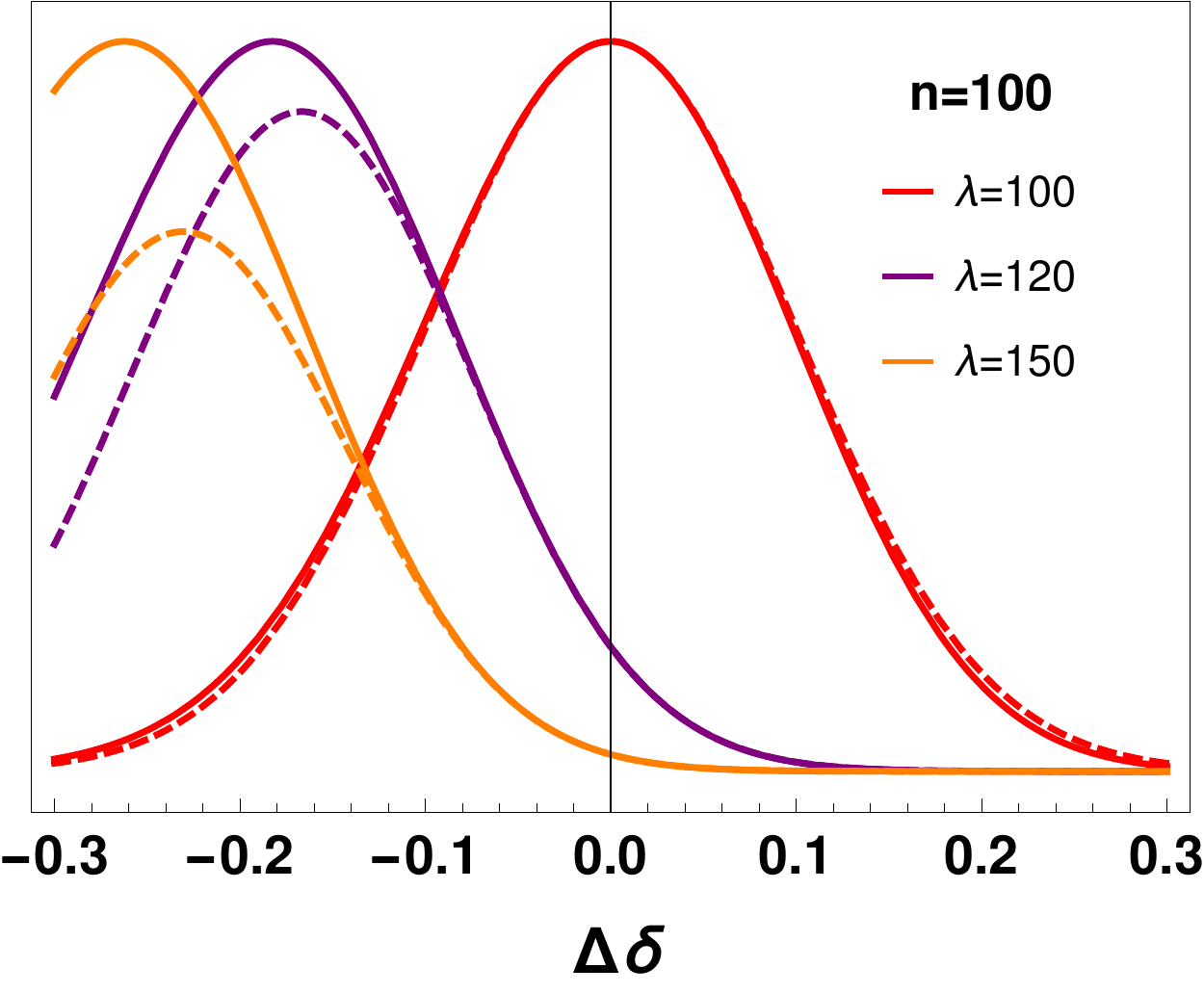}
\includegraphics[scale=0.37,clip=true, trim= 0cm 0cm 0cm 0cm]{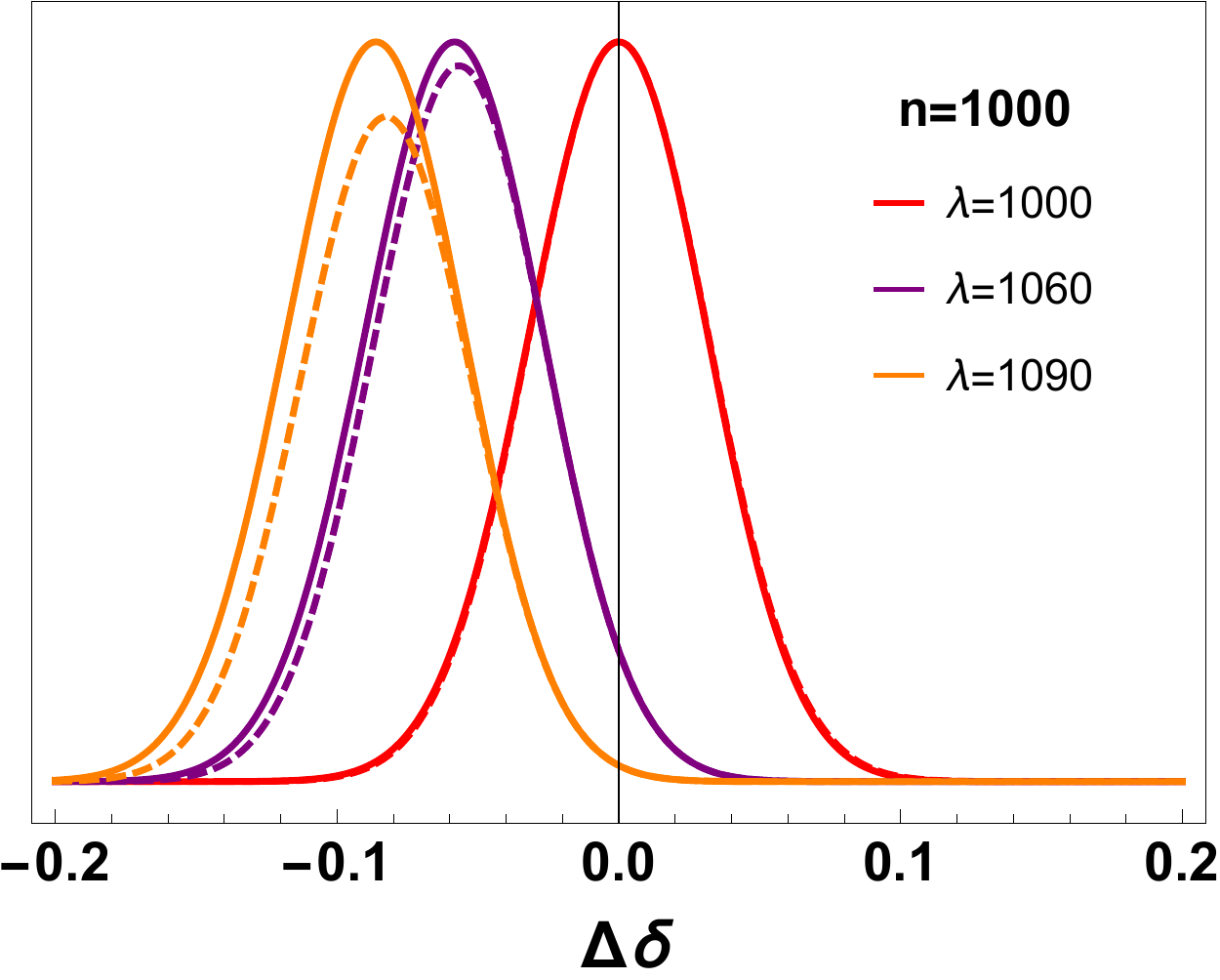}
\caption{
Exact  and approximate Poisson likelihood of Eq.~\eqref{eq:poisapprox} (resp. plain and dashed curves) as a function of the nuisance parameter $\Delta \delta$.
\label{fig:LikeApprox}}
\end{figure}

However, it turns out that one can go further under the extra condition that the \textit{combined} uncertainties are small.  Indeed, one can use a Taylor expansion with respect to the magnitude of the combined uncertainties $\Delta_I$ up to quadratic order in order to simplify the likelihood. This will render possible a completely analytical marginalisation. 
In both cases of  Poisson and Gaussian statistics, the expansion of the likelihood reads
\be
{\rm Pr}(\hat n| \lambda(1+\delta \Delta))={\rm Pr}(\hat n| \lambda) e^{ (\hat n-\lambda)\Delta\delta - \lambda \Delta^2\delta^2/2 +O(\Delta^3) }\,. \label{eq:poisapprox}
\ee
In practice, the validity range of the approximation depends on the amount of data and on the expected number of events.
 This is illustrated in Fig.~\eqref{fig:LikeApprox} for $\hat n=10,100$ and for typical values of $\lambda$ roughly corresponding to $0$, $2$ and $3$ sigma deviations.
 As a very rough rule of thumb  for typical values of $\hat n$ and $\lambda$, one may keep in mind that the validity is good up to $\Delta\sim 20\%$. 
\footnote{In the presence of large sample $\hat n \gg 1$, it is customary to approximate the Poisson likelihood by a Gaussian. It is worth noticing that, while the computation of Eq.~\eqref{eq:poisapprox} is straightforward for the Poisson case, obtaining the same result starting from the Gaussian is a bit more delicate. 
 Depending on how the approximation is done, slightly different expressions can be obtained, that all are close from each other provided that  $\hat n \gg 1$.  The Poisson result Eq.~\eqref{eq:poisapprox} is valid for any $\hat n $, and will be used in the following. }
%\footnote{
%This is because the Gaussian is obtained from the Poisson by performing another quadratic expansion, using the fact that the likelihood forces $\lambda$ to be close from $\hat n$ when $\hat n$ is large. One starts from $\hat n=\lambda(1+\epsilon)$ and expand with respect to to $\epsilon$, which leads to a likelihood going as $\exp(- \lambda \epsilon^2/2)$. 
%The nuisance parameter being itself a small perturbation of $\lambda$,  the correct dependence is obtained by starting with $\hat n=\lambda(1+\epsilon+\delta)$, so that the likelihood has the form $\exp(- \lambda (\epsilon+\delta)^2/2)$.} 

We now plug the approximation Eq.~\eqref{eq:poisapprox} into the general likelihood $\bar L$, given in Eq.~\eqref{eq:Likecomb}. The marginalisation of $\bar L$  is given by 
\be
\tilde L(\theta)=\int d\bar\delta \bar L(\theta,\bar\delta) \bar\pi(\bar\delta)\,,
\label{eq:Lbarmarg}
\ee
where the combined prior $\bar\pi$ is given by Eq.~\eqref{eq:picomb} and involves the correlation matrix $\rho_{IJ}$ of the $\bar\delta$ given in Eq.~\eqref{eq:corr}. The approximate marginal likelihood is found to be
\be
\tilde L (\theta)= L^{\rm stat}(\theta) L^{\rm sys}(\theta) =L^{\rm stat}(\theta)\, \frac{1}{\sqrt{|\eta \rho+1|}} \exp\left(  \frac{1}{2}
 \xi \cdot \Big( \eta  +\rho^{-1}   \Big)^{-1} \cdot  \xi
  \right)\,, \label{eq:Ltildegen}
\ee
where ``$\cdot$'' is matrix multiplication, and one introduced the vector
\be \xi_I=\Delta_{I}\, (\hat n_{I}-\bar n_{0,I})\,,
\ee
the  diagonal matrix $\eta $ with
\be
\eta_{II}=(\Delta_I)^2 \, \bar n_{0,I} \,,
\ee
and 
 $L^{\rm stat}=\prod_I{\rm Pr}(\hat n_{I}|\bar n_{0,I}) $.

Both $\xi$ and $\eta$ depend on the parameters of interest $\theta$ via the expected event numbers  and the combined uncertainties, \ie~one has in general $n_{0,I}(\theta)$, $\Delta_{I}(\theta)$.
  The $L^{\rm stat}$ term is almost the likelihood  with no nuisance parameters, \ie~the piece of  likelihood encoding  the statistical uncertainty, except that it involves the shifted expected event numbers $\bar n_{0,I}$ (see Eq.~\eqref{eq:n0I}). 
 In the limit of zero systematic uncertainty, \ie~$\Delta\rightarrow 0$, $L^{\rm sys}(\theta)$ becomes an irrelevant constant so that $\tilde L (\theta) \rightarrow L^{\rm stat}(\theta)$.
 The $L^{\rm sys}$ encodes most of the effect of the systematic uncertainties. Its effect is to enlarge and shift  the preferred regions of the parameters of interest.

Frequentist marginalisation (\ie~profiling), as well as the Bayesian and frequentist bias methods described in \cite{Fichet:2015xla} can all be treated analytically, substituting Eq.~\eqref{eq:Lbarmarg} by the appropriate operation. 
The approach described above and leading to Eq.~\eqref{eq:Ltildegen} is general. Nevertheless it is  interesting to work out specific cases that are omnipresent in LHC analyses.

% In the limit of zero systematic uncertainty, \ie~$\Delta\rightarrow 0$, $L^{\rm sys}(\theta)$ becomes an irrelevant constant so that $\tilde L (\theta) \rightarrow L^{\rm stat}(\theta)$. \footnote{A slightly less trivial limit is the limit of zero statistical error, \ie~$n_I^{\rm obs}\rightarrow \infty$. In that case a mapping $\delta=\phi(\theta)$ is imposed by the infinitely precise data, so that $L(\theta)$ tends to $\pi^{\rm sys}(\phi(\theta))$ up to an irrelevant constant.}
%
% the $\cdot$ is matrix multiplication and we have introduced

%In this section, building on the previous simplifications, we show that  marginalisation can in fact be carried out analytically in the  cases of Poisson and Gaussian likelihood, which are the most relevant for particle physics and in particular for the Higgs global fits. The neat outcome will be  an analytical likelihood encoding explicitely \textit{all} the systematic uncertainties present in the problem.

\subsection{Signal strength fit}

\label{se:signalstrength}

The general approach summarised by Eq.~\eqref{eq:Ltildegen} can be applied to the very typical case where
the expected number of events is split into a signal and background component, $n=s+b$. The signal can be also further parametrised as $s=\mu s_0$, where the parameter of interest $\mu$ is a ``signal strength modifier'' and $s_0$ is some nominal value for the signal.

In principle both background  and signal are plagued by systematic uncertainties, so that one should distinguish the elementary nuisance parameters for signal and background, $\delta^s$,$\delta^b$.  After a preliminary step of error propagation, the systematic uncertainty on the expected rates take the form
\be
\begin{split}
n_I=&s_I\exp(\Delta^s_{1,I}\cdot\delta^s+(\delta^s)^t\cdot\Delta^s_{2,I}\cdot\delta^s) \\
&+b_I\exp(\Delta^s_{1,I}\cdot\delta^b+(\delta^b)^t\cdot\Delta^b_{2,I}\cdot\delta^b)
\end{split}\,.
\ee
In order to obtain the standard form for propagated errors Eqs.~\eqref{eq:stddelta},~\eqref{eq:stdDelta1},~\eqref{eq:stdDelta2}\,, one defines the overall vector of elementary uncertainties $\delta=(\delta^s,\delta^b)$, and write $n_I=n_{0,I}\exp(\Delta_{1,I}\cdot \delta +\delta^t\cdot\Delta_{2,I}\cdot \delta  )$ where
\be
n_{0,I}=s_{0,I}+b_{0,I}\,,\quad \Delta_{1,I} =\frac{ (s_{0,I}\Delta^s_{1,I},\,b_{0,I}\Delta^b_{1,I})}{s_{0,I}+b_{0,I}}\,,
\ee
\be
(\Delta_{2,I})_{ij} =\frac{1}{{s_{0,I}+b_{0,I}}}\begin{pmatrix}
s_{0,I}(\Delta^s_{2,I}+ (\Delta^s_{1,I})^2/2) & 0  \\
0 & b_{0,I}(\Delta^b_{2,I}+ (\Delta^b_{1,I})^2/2)
\end{pmatrix}-\frac{1}{2}(\Delta_{1,I})_i(\Delta_{1,I})_j
\,.
\ee
This makes contact with the standard notation of Eq.~\eqref{eq:propmult}, and the analytic marginal likelihood is readily given by Eq.~\eqref{eq:Ltildegen}.
If the $N$ independent likelihoods correspond to $N$ measurements of a same process, one has $s_{0,I}=s_0$, $b_{0,I}=b_0$ for every $I$. 
This case will be illustrated in Sec.~\ref{se:example}. 

Here, positive $s$ and $b$ have been assumed. It is also possible to allow  $s$ to take negative values if it is dominated by the destructive interference between the SM and BSM matrix elements. In that case a linear modelisation of the error on $s$ is fine, but one should bear in mind that if the support of the $\delta^s_I$ is such that $b+s$ can be zero, depending on the prior of $\delta^s_I$  the likelihood can blow up above some arbitrary large value of $s$. This is a general fact that is not specific to the approximations studied in this paper.

\subsection{Differential distributions} Another typical measurement at the LHC is the one of a differential distribution. The  likelihood with no systematic uncertainties has then the form of Eq.~\eqref{eq:Likeind}, where every measurement corresponds to a different bin $I$ , and  $\hat n_I$ is the  observed event number in the bin $I$. Denoting by $X\in\mathcal{D}$ the variable along which the events are binned and by $\mathcal{D}_I$ the subdomain of $\mathcal{D}$ defining the bin $I$, the expected event numbers are given by $n_I= n_{\rm tot}\int_{\mathcal{D}_I} dx f_X(x) $.

Differential distributions get deformed by detector effects, that typically smear their shape. A general way of modelling the smearing is
to write the binning variable as
 \be
X=X_0(1+\Delta(X_0)\delta)\,,
\ee
where $\Delta(X_0)$ is the relative magnitude of the uncertainty at the location $X_0$. 
As a simple example, we assume  a model of smearing independent of $X$ -- the general case can be treated similarly. The expected number of events in a bin $I$ is given by
\be
n_I=n_{\rm tot} \int_{\mathcal{D}_I}f_X(x(1+\Delta\delta))dx
\,,
\ee
where $n_{\rm tot}$ is the expected total number of events.
Starting from Eq.~\eqref{eq:Likeind}, one can disentangle the information of shape and total event number,
\be
L(\theta,\delta)=L_{\rm tot}(\theta) L_{\rm shape}(\theta,\delta)\,,\quad
L_{\rm shape}(\theta,\delta)=\prod_{I=1}^n \left(\frac{n_I(\theta,\delta)}{n_{\rm tot}(\theta)}\right)^{\hat n_I}\,.
\ee
Only $L_{\rm shape}$ depends on $\delta$, as this nuisance parameter models a shape deformation. 
Expanding $\log n$ over $\Delta$ at quadratic order gives
\be
n_I=n_{I,0}\exp\left(\Delta_{1,I}\delta+\Delta_2{2,I}\delta  \right)\,,
\ee
with
\be
n_{I,0}=n_{\rm tot}\int_{\mathcal{D}_I} f_X(x)dx\,,\quad
\Delta_{1,I}=\Delta \int_{\mathcal{D}_I}x f'_X(x)dx\left(\int_{\mathcal{D}_I} f_X(x)dx\right)^{-1}\,,
\ee
\be
\Delta_{2,I}=\frac{\Delta^2}{2}\left[\left( \int_{\mathcal{D}_I}x^2 f''_X(x)dx\right)\left(\int_{\mathcal{D}_I} f_X(x)dx\right)^{-1}-\left(\int_{\mathcal{D}_I}x f'_X(x)dx\right)^2\left(\int_{\mathcal{D}_I} f_X(x)dx\right)^{-2}\right]\,.
\ee
As an aside one may notice that when one expands $n_{I}$ at second order,  the quadratic term explicitly shows the effect of smearing. If $f$ is convex (concave) over the bin, then $f_X''>0$ ($f_X''<0$), so that the quadratic term fills (depletes) the bin, accordingly to what is expected from a smearing process.

 Plugging this expression into the likelihood  gives exactly \footnote{Interestingly, no Taylor expansion of the likelihood is needed to get this result.}
\be
L_{\rm shape}=L_{\rm shape}^{\rm stat} e^{\xi\delta-\eta\delta^2/2}\,,
\ee
with  $\xi=\sum_I\Delta_{1,I}$, $\eta=2\sum_I\Delta_{2,I}$.
%\be
%\xi=\Delta \sum_{I=1}^N \frac{\int_{\mathcal{D}_I}xf'_X(x)dx }{\int_{\mathcal{D}_I}f_X(x)dx}\,,\quad
%\eta=\Delta^2\sum_{I=1}^N\left[\left(
%\frac{\int_{\mathcal{D}_I}xf'_X(x)dx }{\int_{\mathcal{D}_I}f_X(x)dx}\right)^2
%-\frac{\int_{\mathcal{D}_I}x^2f''_X(x)dx }{\int_{\mathcal{D}_I}f_X(x)dx}\right]\,.
%\ee
Marginalising with a Gaussian prior for $\delta$ gives once again Eq.~\eqref{eq:Ltildegen}, here in a one-variable version:
\be
\tilde{L}_{\rm shape}=L_{\rm shape}^{\rm stat} \frac{1}{\sqrt{\eta+1}} \exp\left(\frac{ \xi^2}{2(\eta+1)}\right)\,.
\ee

Finally, the unbinned version of the same  likelihood is directly obtained by taking the limit of infinitely thin bins. Then $\hat n_I$ can be only zero or one, the integrals can be simplified and the observed events end up labelled by their position $\hat X_I$. One gets
\be
\Delta_{1,I}=\Delta \hat X_I \frac{f'_X(\hat X_I)}{f_X(\hat X_I)}
\,,\quad \Delta_{2,I}=\frac{\Delta^2}{2} (\hat X_I)^2\left(
\frac{f''_X(\hat X_I)}{f_X(\hat X_I)}-\frac{f'^2_X(\hat X_I)}{f^2_X(\hat X_I)}
\right)\,,
\ee
from which the marginal likelihood follows. 
The information that has to be reported to reconstruct this smeared likelihood  is  
\begin{itemize}
\item The magnitude of the relative uncertainty on the binning variable $X$,
\item The first and second derivatives of the expected shape $f_X$.
\end{itemize}

%
%From a practical point of view, the systematic uncertainties from detector effects (\ie~the smearing) are often unfolded from the observed distribution. 
%However, it is well-known \cite{} that this procedure is fundamentally ill-posed, because a bare unsmearing automatically magnifies the statistical fluctuations present in the sample. In order to deal with this issue \footnote{This is a general issue of so-called ``inverse problems''.}, a
% somewhat arbitrary smoothing function needs to be used, that introduces some degree of arbitrariness in the analysis.
% 
%We emphasize that the analytic method presented here can be used as an alternative to unfolding.
%Indeed, it  
%is another way of transmitting all the  information needed to confront experiment with theory: the detector effect
%
%fulfils the same goal of reproducible 
%

% Rather the analytic method we propose can be used as a way to completely bypass the unfolding procedure.
%

%Provided the expected distribution of events $f_X$ is known, the only information needed is the smearing function $\Delta(x)$ in the unbinned case, or the value of $\Delta$ in every bin in the binned case. Compared to the general list written in Sec.~\ref{se:pract}, the case of a differential distribution is simpler because the partial derivatives are already known the theory side.

\section{An example of signal strength fit }
\label{se:example}

\begin{table}
\center
%\begin{tabular}{|c|c|c|c|}
%\hline
%\multicolumn{4}{|c|}{$\Delta\sim5\%$} \\
%\hline
%Channel & A & B & C \\
%\hline
%$\hat n$ & 280 & 310 & 320 \\
%\hline
%$\Delta^{s,1}$ & $5\%$  & $-5\%$  & $5\%$  \\
%$\Delta^{s,2}$ &  $2.5\%$ & $5\%$  & $5\%$  \\
%$\Delta^{s,3}$ & $0\%$ & $-2.5\%$  & $-2.5\%$ \\
%$\Delta^{b,1}$ & $-5\%$ & $-5\%$ & $5\%$ \\
%$\Delta^{b,2}$ &  $2.5\%$ & $5\%$ & $-5\%$ \\
%\hline
%\end{tabular}\quad\quad
\begin{tabular}{|c|c|c|c|}
\hline
%\multicolumn{4}{|c|}{$\Delta\sim10\%$} \\
%\hline
Channel & A & B & C \\
\hline
$\hat n$ & 280 & 310 & 320 \\
\hline
$\Delta^{s,1}$ & $10\%$  & $-10\%$  & $10\%$  \\
$\Delta^{s,2}$ &  $5\%$ & $10\%$  & $10\%$  \\
$\Delta^{s,3}$ & $0\%$ & $-5\%$  & $-5\%$ \\
$\Delta^{b,1}$ & $-10\%$ & $-10\%$ & $10\%$ \\
$\Delta^{b,2}$ &  $5\%$ & $10\%$ & $-10\%$ \\
\hline
\end{tabular}
\caption{
Observed data and nuisance parameters in three statistically independent channels $A,B,C$. The average expected background is taken to be $b_0=100$.
\label{tab:data} }
\end{table}

\begin{figure}[t]
\begin{picture}(400,200)
\put(00,0){\includegraphics[scale=0.55,clip=true, trim= 0cm 0cm 0cm 0cm]{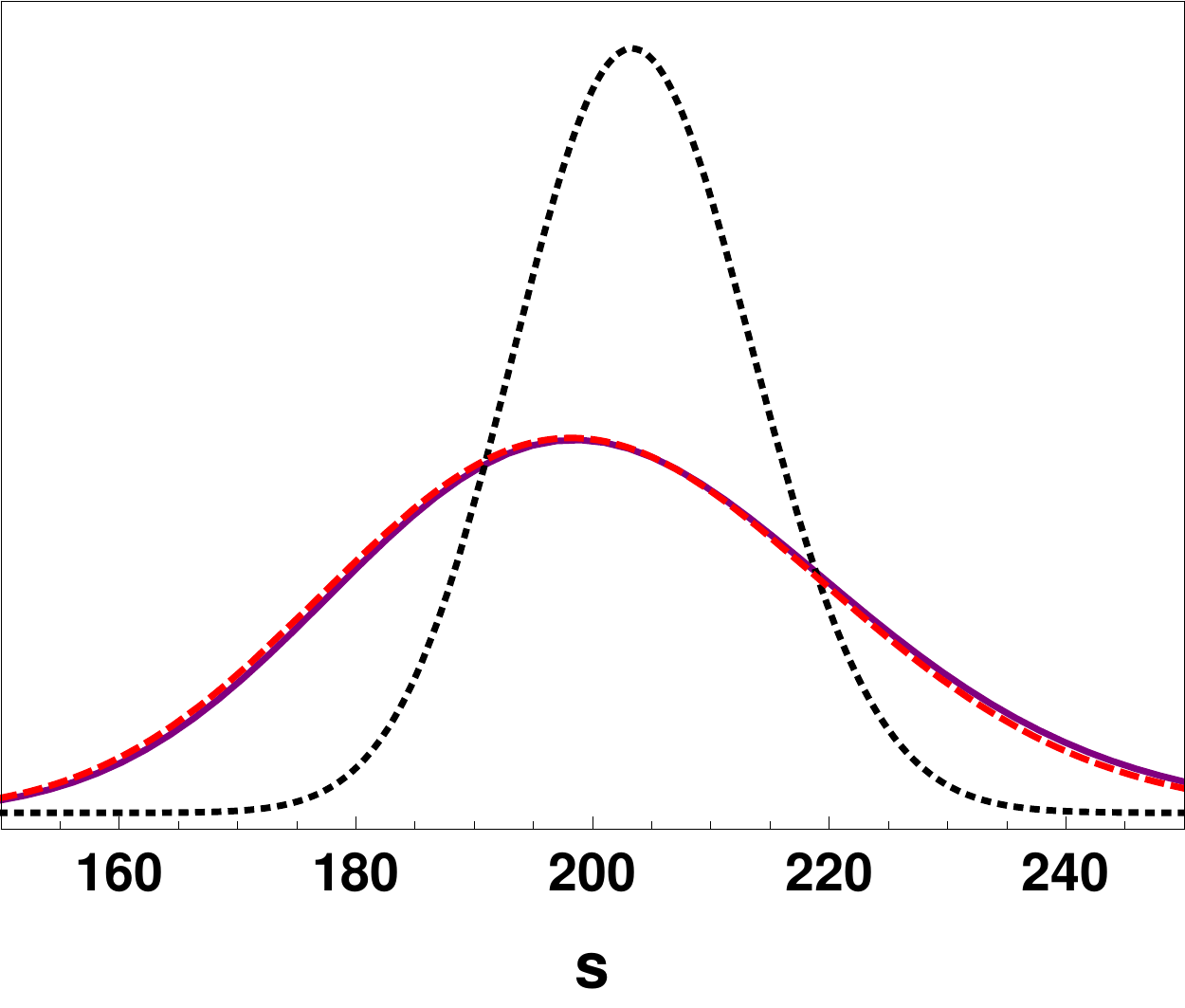}}
\put(220,0){\includegraphics[scale=0.55,clip=true, trim= 0cm 0cm 0cm 0cm]{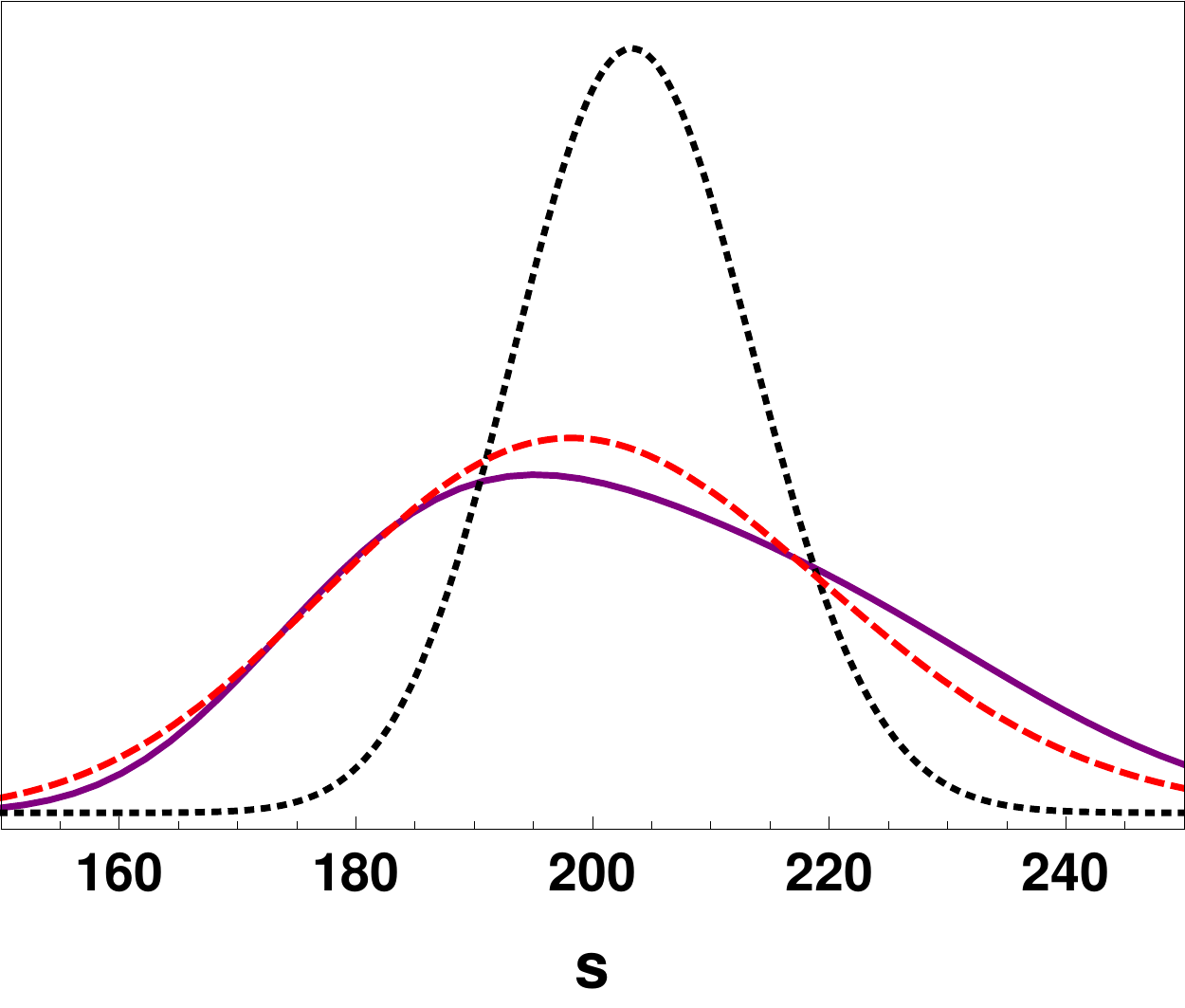}}
\put(20,120){{\color{blue} $\Delta\sim10\%$}}
\put(5,110){{\color{blue} \textit{-- log-normal --}}}
\put(240,120){{\color{blue} $\Delta\sim10\%$}}
\put(225,110){{\color{blue} \textit{-- log-flat --}}}
\put(122,130){{\color{black} $L^{\rm stat}(s)$}}
\put(342,130){{\color{black} $L^{\rm stat}(s)$}}
\put(140,80){{\color{red} $\tilde L(s)$}}
\put(360,80){{\color{red} $\tilde L(s)$}}
%Quadrature / no correlation
\end{picture}
\caption{ Likelihood functions drawn from  data of Tab.~\ref{tab:data}.
Plain line: Exact marginal likelihood, evaluated numerically.
Dashed line: Approximate marginal likelihood, evaluated analytically.
Dotted line: likelihood with no systematic uncertainties, $L^{\rm stat}(s)\equiv L(s,\delta=0)$.
\label{fig:LikeMarg}}
\end{figure}

In order to illustrate our results,  we consider a somewhat realistic scenario  for the charac\-terisation  of a  signal. Formally, the scenario considered corresponds to a particular case of  the signal strength analysis  described in Sec.~\ref{se:signalstrength}. This example will also be used to check the accuracy  of the approximate marginal likelihood, $\tilde L$.

To carry out this toy analysis one first has to setup the ``observed'' data, the expected background and the systematic uncertainties.
We assume  three independent observation channels $I=(A,B,C)$. An observed event number  $\hat n_{A,B,C}$ is assumed for each channel. The expected event number is given a signal+background form  $n=s+b$, which is taken to be common to all channels, so that $s_I=s$, $b_I=b$, and the  nominal value of $b$ is fixed. 

We further assume the presence of 3 independent systematic uncertainties labelled $a,b,c$ for the signal and 2 uncertainties labelled $d,e$ for the background. Both signal and background are positive, so that we use the error modelisation of Eq.~\eqref{eq:deltaexp}. We are going to consider cases of a flat and Gaussian prior for $\delta$, which imply respectively log-normal and log-flat priors for the $s$, $b$ components of the expected event number.

 Assuming a first step of propagation from more elementary uncertainties, and disregarding the possible $\Delta_2$ terms for simplicity, the leading effect of the systematic uncertainties on the expected event number is characterised by $3\times 5$ numbers $\Delta_{1,I}^{s/b,i}$, given in Tab.~\ref{tab:data}. In the notation of Sec.~\ref{se:pract}, one has $N=3$, $p=5$. Note that with this starting point,  the magnitude of the $p$ elementary uncertainties are already combined with the $N\times p$ derivatives of the  expected signal. 
The uncertainties appear in each  channel $I$  as $s=s_{0}\exp(\Delta^{s}_{1,I}.\delta^s)$, $b=b_{0}\exp(\Delta^b_{1,I}.\delta^b)$.

Having characterised the main effect of the systematic uncertainties, we can readily use the approximate marginal likelihood $\tilde L$. Moreover, we  compute  the exact marginal likelihood by numerically integrating over the  five nuisance parameters, \footnote{Note that for a flat distribution, ${\rm V}(\delta)=1$ implies  $\delta\in[-\sqrt{3}, \sqrt{3}]$.} 
which provides a way of testing directly the accuracy of $\tilde L$.
 In practice,  plotting the exact marginal likelihood of our example takes about an hour on a laptop with average specifications, while plotting the approximate likelihood is instantaneous.

All the  numbers assumed for $\hat n_I$ and the $\Delta_{1,I}^{s/b,i}$ are given in Tab.~\ref{tab:data}. 
The observed numbers are chosen so that the statistical uncertainty be of $O(10\%)$. This is for example the case in the global fit of the 8 TeV Higgs signal strengths.
The signs of the systematic uncertainties have been chosen  so that the likelihood strongly depends on every nuisance parameter. The relative magnitudes of the elementary uncertainties are chosen to be  $O(10\%)$.

This illustrative scenario can be used  in order to check the accuracy of the two kinds of approximations required to obtain the $\tilde L$ likelihood: \textit{a)} the CLT-based Gaussian approximation (see Sec.~\ref{se:taming}) and \textit{b)} the likelihood expansion (see Sec.~\ref{se:anal}).
%~\footnote{In this example the uncertatines considered are already propagated into the event numbers, hence approximation \textit{a)} corresponds directly to the convergence rate of the CLT.}
 The Gaussian and flat priors allow us to disentangle between these two approximations, because for  Gaussian priors, approximation \textit{a)} is always satisfied, \ie~the CLT is perfectly convergent. Thus for Gaussian priors,  the  discrepancies between the approximate and local likelihoods come only from approximation \textit{b)}. In contrast, in the flat prior case the discrepancies come from both approximations \textit{a)} and \textit{b)}.

The exact and approximate marginal likelihoods are shown in Fig.~\ref{fig:LikeMarg}. In the case of Gaussian priors for $\delta$ (\ie~log-normal uncertainties), the two curves agree very well. This shows that  approximation \textit{b)} is well under control. 
In order to test approximation \textit{a)}, \ie~the CLT convergence,  we can now compare with  the case of  flat priors for $\delta$. It turns out that a mild discrepancy appears in certain regions. 
This illustrates the degree of convergence of the CLT in a five-parameters case. The two curves agree still fairly well in this flat-prior case, in the sense that the best-fit regions  drawn from these likelihoods would be  similar.
For a larger number of nuisance parameters, this discrepancy is expected to decrease as the CLT convergence should improve.

%
%In the $\Delta\sim5\%$ case, the two curves agree very well, in the sense that the best-fit regions  drawn from these likelihoods would be essentially identical. This case is rather encouraging as it shows that all the approximations made to obtain the analytical formula of $\tilde L(s)$ are well under control. In particular, even though there are only five nuisance parameter with flat prior,  the central limit theorem is at work with an already good degree of convergence. 
%
%In order to explore the validity of our approximations, we also consider a case where magnitudes are twice larger. Here  a somewhat larger discrepancy appears, \footnote{The shift of the preferred region occurs because there is some tension between the observed values, which is modified in the presence of the nuisance parameters (see also \cite{Fichet:2015xla}).}  which occurs at  the level of the expansion Eq.~\eqref{eq:poisapprox} (see also Fig.~\ref{fig:LikeApprox}). 
%
%The origin of the discrepancy originates partly from the expansion and partly from the CLT approximation (the latter can easily be checked by replacing the flat elementary priors by Gaussian ones). Also the discrepancy \textbf{[... if one change the signs ]}
%
%Even though  the discrepancy is more obvious in this case,  it still should be contrasted with the original likelihood with no systematic uncertainties (dotted line on the plots). We conclude that even this case  gives a fair account of the effect of the systematic uncertainties.

\section{Conclusion}

With the goal of simplifying the treatment of systematic uncertainties in typical LHC analyses, 
we have studied the behaviour of a generic likelihood in the presence of a large number of uncertainties with small relative magnitudes. 

Whenever this  condition  is satisfied, it turns out  that well-controled approximations become available, which provide a way of drastically simplifying the incorporation of systematic uncertainties into the likelihood. Our demonstration is split into  steps of error propagation and  error combination. In  the latter, the Lyapunov central limit theorem applies to the combined uncertainties, thereby approximating their joint distribution as a multivariate normal.
%The Lyapunov condition is easily met any of the common LHC nuisance parameter.
 This implies that the shape of the priors of the elementary uncertainties is irrelevant -- only their magnitudes matter.

Whenever the combined uncertainties are small enough, say $\lesssim 20\%$, the likelihood can be further simplified and the complete marginal likelihood is obtained analytically. This general result is applied to the important cases of signal strength characterisation  and  differential distribution smearing.

For illustration, we present a toy-analysis of signal strength characterisation including  systematic uncertainties on signal and background. 
The approximate and  exact marginal likelihoods are found to be in fairly good agreement in this example, implying that all approximations are well under control.

Beyond the obvious  gain of avoiding heavy numerical marginalisation, another practical matter is the communication of systematic uncertainties, for example from an experiment to the public. Our approach implies that all the needed information is encoded  into a finite set of numbers, namely the relative magnitude of  elementary uncertainties and the derivatives of the expected event numbers. The transmission of this information is straightforward, and gives a fairly human-readable summary of the systematic uncertainties. In principle, this simple method  could be used to make public the detector effects that are included in LHC analyses.

The marginal likelihood presented in this paper is purely Bayesian.  It is also  possible to compute analytically the marginal likelihood in case of a frequentist profiling (described in App.~\ref{app:freq}), as well as to apply the bias methods formalised in Ref.~\cite{Fichet:2015xla}. 
Finally, although our study is oriented towards LHC analyses, it could also be  readily applied into other experimental contexts.

\section*{Acknowledgements} 

I would like to thank G. Moreau, S. Kraml, E. Ponton, G. von Gersdorff for useful discussions and comments.
 This work is supported by the
S\~ao Paulo Research Fundation (FAPESP) under grant 2014/21477-2.
\\
\\
\\

\appendix

\section{Analytic frequentist marginalisation}
\label{app:freq}
In Sec.~\ref{se:anal}, a frequentist marginalisation of the combined uncertainties can also be done. This is obtained by substituting Eq.~\eqref{eq:Lbarmarg} with 
\be
\tilde L(\theta){\rm freq}=\max_{\bar \delta} \left[ \bar L(\theta,\bar\delta) \bar\pi(\bar\delta) \right]\,.
\ee
The resulting approximate likelihood is
\be
\tilde L (\theta){\rm freq}= L^{\rm stat}(\theta)\,  \exp\left(  \frac{1}{2}
 \xi \cdot \Big( \eta  +\rho^{-1}   \Big)^{-1} \cdot  \xi
  \right)\,, \label{eq:Ltildegenfreq}
\ee
which ressembles very much the Bayesian result up to a $\sqrt{|\eta\rho+1|}$ factor. Typically, the variation of this factor with respect to the parameters of interest is small compared to the variation of the exponential term. Hence, in practice,   the frequentist and Bayesian approximate likelihoods are almost equivalent. The subsequent frequentist and Bayesian best-fit regions obtained from these  likelihoods  thus differ mostly by the definition of frequentist and Bayesian contours~\cite{Agashe:2014kda}.

%\end{thebibliography}

\end{document}